\documentclass[dvipsnames]{article}
\usepackage{spconf,amsmath,graphicx,hyperref}
\usepackage{etoolbox}
\usepackage{booktabs}

\usepackage{booktabs,multirow}
\usepackage{xcolor}
\usepackage{float} 

\usepackage{pifont} 
\newcommand{\cmark}{\textcolor{ForestGreen}{\ding{51}}} 
\newcommand{\xmark}{\textcolor{BrickRed}{\ding{55}}}    

\title{Leveraging Large Multimodal Models for Audio-Video Deepfake Detection: A Pilot Study}
\name{Songjun Cao$^{1,\dagger}$ \qquad Yuqi Li$^{1,2,\dagger}$ \qquad Yunpeng Luo$^{1}$ \qquad Jianjun Yin$^{2}$ \qquad Long Ma$^{1}$}

\address{$^{1}$ Tencent Youtu Lab, China \\
    $^{2}$ Fudan University, China \\
    \vspace{-1.5em} 
    \textsuperscript{$^{\dagger}$}These authors contributed equally to this work. 
   }

\begin{document}

\maketitle
\setlength{\parskip}{0pt} 
\setlength{\abovedisplayskip}{0pt} 
\setlength{\belowdisplayskip}{0pt}

\begin{abstract}
Audio-visual deepfake detection (AVD) is increasingly important as modern generators can fabricate convincing speech and video. Most current multimodal detectors are small, task-specific models: they work well on curated tests but scale poorly and generalize weakly across domains. We introduce \textbf{\textit{AV-LMM\-Detect}}, a supervised fine-tuned(SFT)  large multimodal model that casts AVD as a prompted yes/no classification—``Is this video real or fake?". Built on Qwen 2.5 Omni,  it jointly analyzes audio and visual streams for deepfake detection and is trained in two stages: lightweight LoRA alignment followed by audio-visual encoder full fine-tuning. On FakeAVCeleb and Mavos-DD, \textbf{\textit{AV-LMM\-Detect}} matches or surpasses prior methods and sets a new state of the art on Mavos-DD datasets.
\end{abstract}

\begin{keywords}
audio-visual deepfake detection, large multimodal models, fine-tuning, cross-modal forensics
\end{keywords}

\section{Introduction}
The realism of synthetic media has surged with recent generative models, making audio-visual deepfake detection (AVD) a pressing problem for media integrity and public trust. Early research concentrated on \emph{visual-only} detectors, including 
CNN/geometry pipelines such as MesoNet~\cite{afchar2018mesonet}, Capsule\cite{nguyen2019capsule}, Xception-based baselines\cite{chollet2017xception}, and more recent designs like LipForensics\cite{haliassos2021lips}, Multiple-Attention\cite{zhao2021multi}, and SLADD\cite{chen2022self}. While effective on specific manipulations, vision-only systems are intrinsically blind to cross-modal inconsistencies and often degrade under distribution shift.

To address this, \emph{audio-visual} (A-V) methods jointly model speech and video, e.g. AVN-J\cite{qian2021audio}, Emotion Don't Lie\cite{mittal2020emotions}, AVFakeNet\cite{ilyas2023avfakenet}, VFD\cite{cheng2023voice}, AVoID-DF\cite{yang2023avoid}, and AVFF
\cite{oorloff2024avff}. These systems generally improve robustness by aligning lip motion, acoustic content, and facial dynamics. However, most pipelines remain task-specific and relatively small, which limits scalability, adaptation to novel forgeries, and cross-domain generalization observed in large-scale deployments. 

\begin{figure}[t!]
    \centering
    \includegraphics[width=0.5\textwidth, trim=0 140 450 0, clip]{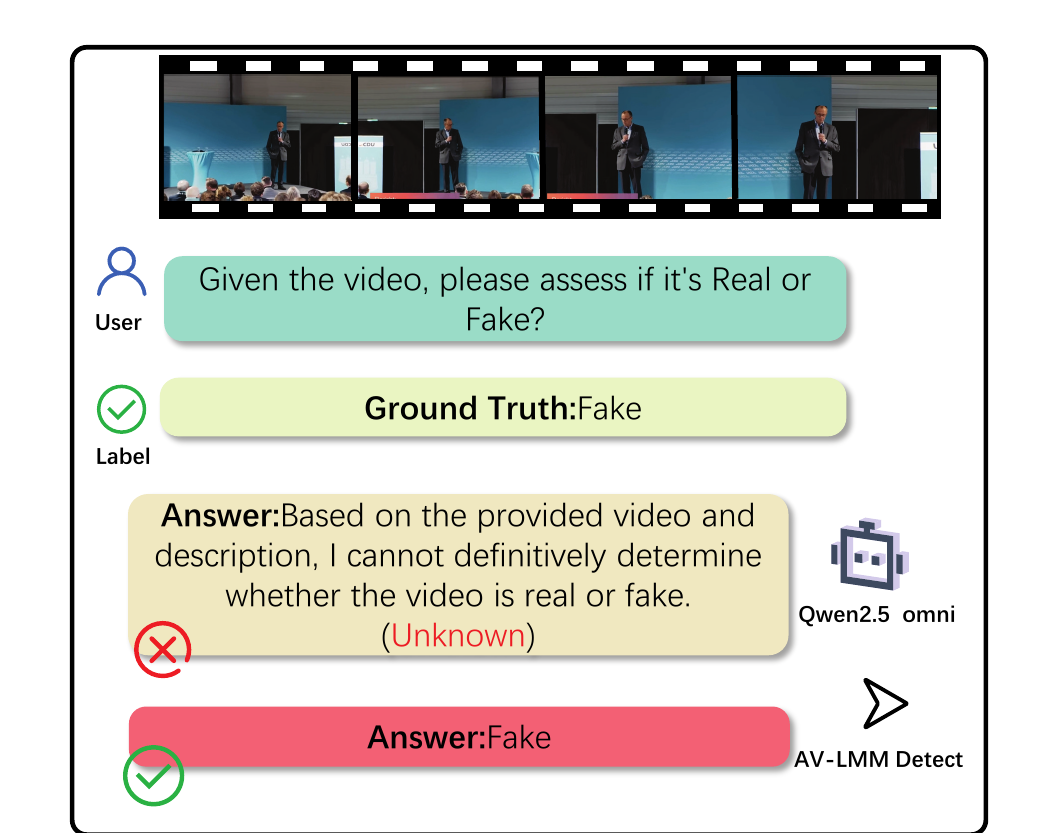}
    \caption{Performance comparison with Qwen 2.5 Omni~\cite{xu2025qwen2}, highlighting the improved results of our AV-LMMDETECT. The video data is from the open-source MAVOS-DD dataset\cite{croitoru2025mavos}.}
    \label{fig:performance_comparison}
\end{figure}

In parallel, the \emph{audio-only} LLM line—exemplified by \textbf{ALLM4ADD}\cite{gu2025mathcal}—has shown that large (audio) language models are promising for deepfake detection when appropriately reformulated and fine-tuned, but purely audio systems cannot exploit visual cues and struggle with audio-video mismatch.

We explore whether an \emph{supervised fine-tuned (SFT) large multimodal model (LMM)} can serve as a \emph{unified} AVD detector. We present \textbf{\textit{AV-LMMDetect}}, which casts AVD as a prompted binary question—``\emph{Is this video real or fake?}''—and fine-tunes Qwen 2.5 Omni~\cite{xu2025qwen2} to \emph{jointly} judge over audio and visual streams. 

As illustrated in Fig.~\ref{fig:performance_comparison}, our \textbf{\textit{AV-LMMDetect}} demonstrates superior detection capability compared to the base Qwen 2.5 Omni~\cite{xu2025qwen2} model. While the base model responds with uncertainty content, our fine-tuned approach correctly identifies the deepfake video as ``Fake'', validating the effectiveness of our specialized SFT fine-tuning for audio-visual deepfake detection tasks. The quantitative comparison in Table~\ref{tab:ft-open-set-table} shows our fine-tuned approach achieves 85.09\% accuracy versus 32.26\% for the base model in the most challenging open-set full scenario.

A two-stage regimen is adopted: (i) lightweight LoRA alignment for efficient adaptation, followed by (ii) audio-visual encoder full fine-tuning to maximize cross-modal synergy.

\noindent\textbf{Contributions.}
\begin{itemize}
  \item We introduce \textbf{\textit{AV-LMMDetect}}, the first supervised fine-tuned (SFT) large multimodal model for end-to-end audio-visual deepfake detection through prompted classification.
  \item We propose a two-stage training strategy (LoRA alignment $\rightarrow$ audio-visual encoder full tuning) that preserves efficiency while achieving strong cross-modal performance.
  \item On \textbf{FakeAVCeleb}\cite{khalid2021fakeavceleb} and \textbf{Mavos-DD}\cite{croitoru2025mavos}, \textbf{\textit{AV-LMM\-Detect}} achieves competitive performance, with state-of-the-art results on MAVOS-DD and comparable performance to SOTA methods on FakeAVCeleb.
\end{itemize}

\noindent Overall, these results indicate that supervised fine-tuned (SFT) LMMs are a viable path toward robust, generalizable AVD, complementing and extending prior small-model pipelines and audio-centric LLM efforts.

\section{Related Work}
\label{sec:format}

\textbf{Large Language Models for Media Forensics.} Recent advances in large language models have shown promising applications in media forensics. ALLM4ADD~\cite{gu2025mathcal} pioneered the use of audio large language models for deepfake detection by reformulating the task as audio question answering. However, these approaches are limited to single modalities and cannot exploit cross-modal inconsistencies that are crucial for robust detection.

\textbf{Multimodal Fusion Strategies.} Traditional audio-visual detection methods employ various fusion approaches, including late fusion~\cite{qian2021audio}, and attention-based mechanisms~\cite{ilyas2023avfakenet}. While effective, these approaches typically rely on hand-crafted fusion architectures and struggle with generalization across diverse manipulation techniques and domains.

\textbf{Instruction Tuning for Multimodal Tasks.} The success of instruction tuning in natural language processing has inspired its application to multimodal understanding tasks. Recent works have demonstrated that supervised fine-tuned (SFT) models can achieve strong performance on vision-language tasks~\cite{xu2025qwen2}. Our work extends this paradigm to the forensics domain, leveraging the reasoning capabilities of large multimodal models for deepfake detection.

\section{Methodology}
\label{sec:pagestyle}

\subsection{Two-Stage Training Strategy}

\begin{figure}[htbp]
    \centering
    \includegraphics[width=0.6\textwidth, trim=0 230 180 0, clip]{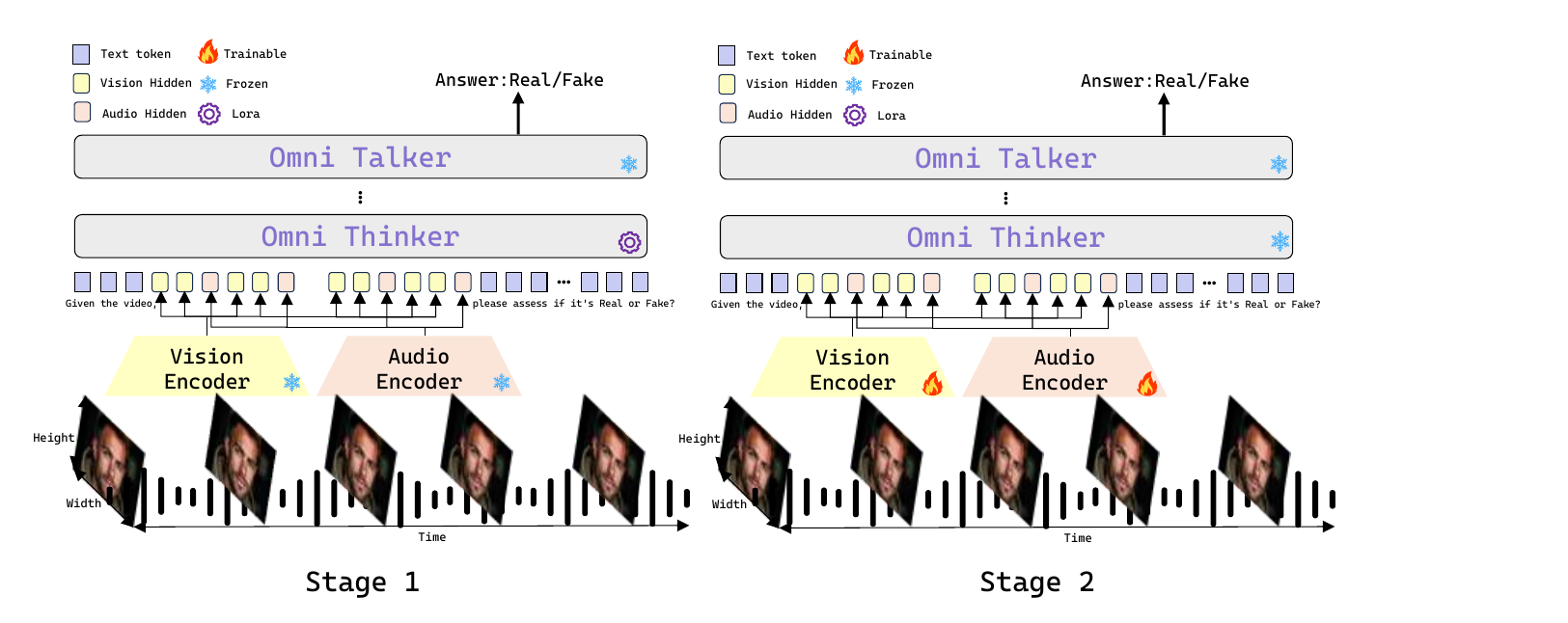}
    \caption{Overview of our \textbf{\textit{AV-LMMDetect}}. We reformulate audio-visual deepfake detection as a multimodal question answering task. During two-stage fine-tuning, Stage 1 employs LoRA for efficient alignment, while Stage 2 opens vision and audio encoders for audio-visual encoder full learning. The video data is from the open-source MAVOS-DD dataset\cite{croitoru2025mavos}.}
    \label{fig:methodology_overview}
\end{figure}

Our training approach consists of two complementary stages illustarted in Fig.~\ref{fig:methodology_overview}:

\textbf{Stage 1: LoRA Alignment.} We perform lightweight LoRA fine-tuning with frozen vision and audio encoders. This stage focuses on aligning the language model's reasoning capabilities with the deepfake detection task through specific instructions: ``Only answer `Real' or `Fake''', ensuring efficient adaptation while preserving the model's general knowledge.

\textbf{Stage 2: Audio-Visual Encoder Full Fine-tuning.} We unlock the vision and audio encoders and conduct full fine-tuning across all modalities. This stage maximizes cross-modal synergy by allowing the model to learn task-specific multimodal representations and capture subtle audio-visual inconsistencies indicative of deepfake manipulation. 

We also experimented with reversing the order of Stage 1 and Stage 2, and found that the results remained essentially consistent.

\subsection{Question Answering Formulation}
We reformulate audio-visual deepfake detection as a binary question answering task. Given an input video with audio, the model is prompted with: ``\textit{Given the video, please assess if it's Real or Fake?}'' The model generates responses from a constrained vocabulary containing only ``Real'' and ``Fake'' tokens (token IDs: 12768 and 52317, respectively). 

We construct a fine-tuning dataset $\mathcal{D}_{ft}$ by pairing each audio-visual input with the corresponding prompt instruction. The training objective is to minimize the language modeling loss function $\mathcal{L}$ over $\mathcal{D}_{ft}$:

\begin{equation}
\theta^* = \arg \min_{\theta} \sum_{i=1}^{N} \mathcal{L}(\mathcal{M}_{\theta}(V^i, A^i, q), y^i),
\end{equation}
where $\theta$ represents the trainable parameters of the model $\mathcal{M}$, $V^i$ and $A^i$ are the visual and audio inputs respectively, $q$ is the prompt question, $y^i$ is the ground truth label (``Real'' or ``Fake''), and $N$ is the number of training samples. These token IDs correspond to the model's vocabulary, facilitating the extraction of prediction logits for subsequent evaluation and analysis.

\subsection{Top-2 Token Prediction and Evaluation}
We select the two tokens from the model's vocabulary output, which correspond to our target ``Real'' and ``Fake'' tokens (token IDs: 12768 and 52317). We then use the corresponding probabilities of ``Real'' and ``Fake'' tokens for classification.

\textbf{Evaluation Metrics:} We employ standard binary classification metrics:
\begin{itemize}
    \item \textbf{Accuracy (Acc):} Proportion of correct predictions
    \item \textbf{Area Under Curve (AUC):} Area under the ROC curve, measuring the model's ability to distinguish between classes across all threshold settings
    \item \textbf{Mean Average Precision (mAP):} Average precision across different recall levels, particularly important for imbalanced datasets
\end{itemize}

The probabilistic output enables robust evaluation across different decision thresholds, providing comprehensive performance assessment beyond simple accuracy metrics.

\section{Experiment}
\label{sec:typestyle}
\subsection{Datasets}

\begin{itemize}
    \item \textbf{FakeAVCeleb\cite{khalid2021fakeavceleb}}: English-language audio-visual deepfakes with face swapping and reenactment methods. We use 70\% for fine-tuning and 30\% for evaluation.
    
    \item \textbf{MAVOS-DD\cite{croitoru2025mavos}}: Multilingual dataset with 250+ hours of real/fake videos in eight languages, featuring seven deepfake methods and four evaluation splits: \textit{In-domain}, \textit{Open-set language}, \textit{Open-set model}, and \textit{Open-set full}.
\end{itemize}

\subsection{Experiment Results}

We evaluate \textbf{\textit{AV-LMMDetect}} on two challenging benchmarks: FakeAVCeleb\cite{khalid2021fakeavceleb} and MAVOS-DD\cite{croitoru2025mavos} , comparing against state-of-the-art vision-only and audio-visual methods. Our experimental results demonstrate the effectiveness of our supervised fine-tuned (SFT) multimodal approach across different evaluation scenarios.

\begin{table}[H]
\centering
\setlength{\tabcolsep}{7pt}
\renewcommand{\arraystretch}{1.1}
\begin{tabular}{lccc}
\toprule
\textbf{Method} & \textbf{Modality} & \textbf{AUC} & \textbf{acc} \\
\midrule
MesoNet\cite{afchar2018mesonet}            & \textcolor{red}{V}   & 60.9 & 57.3 \\
Capsule\cite{nguyen2019capsule}            & \textcolor{red}{V}   & 70.9 & 68.8 \\
Xception\cite{chollet2017xception}           & \textcolor{red}{V}   & 70.5 & 67.9 \\
LipForensics\cite{haliassos2021lips}       & \textcolor{red}{V}   & 82.4 & 80.2 \\
Multiple-Attention\cite{zhao2021multi} & \textcolor{red}{V}   & 79.3 & 77.6 \\
SLADD\cite{chen2022self}              & \textcolor{red}{V}   & 72.1 & 70.5 \\
\midrule
AVN-J\cite{qian2021audio}              & \textcolor{green!60!black}{A-V} & 77.6 & 73.2 \\
Emotion Don't Lie\cite{mittal2020emotions}  & \textcolor{green!60!black}{A-V} & 79.8 & 78.1 \\
AVFakeNet\cite{ilyas2023avfakenet}          & \textcolor{green!60!black}{A-V} & 83.4 & 78.4 \\
VFD\cite{cheng2023voice}                & \textcolor{green!60!black}{A-V} & 86.1 & 81.5 \\
AVoiD-DF\cite{yang2023avoid}           & \textcolor{green!60!black}{A-V} & 89.2 & 83.7 \\
AVFF\cite{oorloff2024avff}               & \textcolor{green!60!black}{A-V} & \underline{99.1} & \textbf{98.6} \\
\midrule
\textbf{AV-LMMDETECT (Ours)} & \textbf{\textcolor{green!60!black}{A-V}} & \textbf{99.2} & \underline{98.02} \\
\bottomrule
\end{tabular}
\caption{Intra-manipulation evaluation on FakeAVCeleb\cite{khalid2021fakeavceleb}. Following standard protocol, we use 70\% as training set and 30\% as test set. We report Area Under the Curve (AUC) and Accuracy (Acc) scores (\%).}
\label{tab:avd_results}
\end{table}

\textbf{Performance on FakeAVCeleb.} Table~\ref{tab:avd_results} shows our \textbf{\textit{AV-LMMDetect}} achieves 98.02\% accuracy and 99.2\% AUC, achieving comparable performance to the current SOTA method AVFF (98.6\% accuracy, 99.1\% AUC). Our approach significantly outperforms traditional vision-only methods (MesoNet: 57.3\%, Xception: 67.9\%) and most audio-visual baselines (AVN-J: 73.2\%, VFD: 81.5\%), demonstrating the effectiveness of supervised fine-tuned (SFT) large multimodal models for audio-visual deepfake detection.

\textbf{Performance on MAVOS-DD.} Table~\ref{tab:ft-open-set-table} shows our model's performance under the official MAVOS-DD\cite{croitoru2025mavos} evaluation protocol, which tests generalization across four challenging scenarios: in-domain, open-set model, open-set language, and open-set full. Our Qwen 2.5 Omni~\cite{xu2025qwen2}-based approach achieves state-of-the-art results in three out of four scenarios, with particularly strong performance in the most challenging open-set full scenario (mAP: 0.96, AUC: 0.92, Acc: 85.09\%).

\begin{table*}[t]
\small
\setlength{\tabcolsep}{5pt}
\renewcommand{\arraystretch}{1.15}
\begin{tabular}{l c ccc ccc ccc ccc}
\toprule
\multirow{2}{*}{\textbf{Method}} & \multirow{2}{*}{\textbf{Fine-tuned}}
 & \multicolumn{3}{c}{\textbf{In-domain}}
 & \multicolumn{3}{c}{\textbf{Open-set model}}
 & \multicolumn{3}{c}{\textbf{Open-set language}}
 & \multicolumn{3}{c}{\textbf{Open-set full}} \\
\cmidrule(lr){3-5} \cmidrule(lr){6-8} \cmidrule(lr){9-11} \cmidrule(lr){12-14}
 & & mAP & AUC & acc & mAP & AUC & acc & mAP & AUC & acc & mAP & AUC & acc \\
\midrule
AVFF \cite{oorloff2024avff} & \xmark & 0.51 & 0.51 & 52.45 & 0.50 & 0.50 & 22.58 & 0.51 & 0.51 & 59.46 & 0.50 & 0.50 & 35.34 \\
MRDF \cite{zou2024cross} & \xmark & 0.50 & 0.46 & 44.04 & 0.52 & 0.52 & 58.04 & 0.46 & 0.41 & 39.35 & 0.51 & 0.49 & 50.78 \\
TALL \cite{xu2023tall} & \xmark & 0.49 & 0.48 & 50.74 & 0.50 & 0.51 & 39.22 & 0.48 & 0.47 & 50.78 & 0.50 & 0.49 & 44.63 \\
\textbf{Qwen 2.5 Omni} \cite{xu2025qwen2} & \xmark & 0.47 & 0.44 & 49.25 & 0.77 & 0.45 & 20.84 & 0.38 & 0.43 & 55.50 & 0.61 & 0.41 & 32.26 \\
\midrule
AVFF \cite{oorloff2024avff} & \cmark & \underline{0.95} & \underline{0.95} & \underline{86.93}
                         & \underline{0.85} & \underline{0.89} & 75.34
                         & \textbf{0.90} & \textbf{0.90} & \underline{84.26}
                         & \underline{0.87} & \underline{0.89} & 77.68 \\
MRDF \cite{zou2024cross} & \cmark & 0.90 & 0.90 & 84.27
                         & 0.78 & 0.88 & \underline{78.32}
                         & \underline{0.88} & \underline{0.88} & 82.15
                         & 0.82 & 0.86 & \underline{78.87} \\
TALL \cite{xu2023tall} & \cmark & 0.87 & 0.86 & 78.07
                         & 0.79 & 0.84 & 66.20
                         & 0.80 & 0.80 & 73.25
                         & 0.77 & 0.79 & 67.42 \\
\midrule
\textbf{AV-LMMDETECT (Ours)} & \cmark
                              & \textbf{0.97} & \textbf{0.97} & \textbf{92.92}
                              & \textbf{0.98} & \textbf{0.94} & \textbf{87.91}
                              & \underline{0.88} & \textbf{0.90} & \textbf{85.58}
                              & \textbf{0.96} & \textbf{0.92} & \textbf{85.09} \\
\bottomrule
\end{tabular}
\caption{Comparison on MAVOS-DD\cite{croitoru2025mavos}. Best results are in \textbf{bold}; second-best are \underline{underlined}.}

\label{tab:ft-open-set-table}
\end{table*}

The results reveal several key insights: (1) Fine-tuning is crucial for all methods, as evidenced by the dramatic performance improvement compared to non-fine-tuned baselines. (2) Our approach demonstrates superior generalization capability, maintaining relatively stable performance across different open-set scenarios 
. (3) The supervised fine-tuned (SFT) formulation enables better cross-modal reasoning, particularly beneficial when encountering unseen generative models.

\subsection{Ablation Experiment}

To validate our training strategy, we conduct ablation studies on MAVOS-DD\cite{croitoru2025mavos} Open-set full scenario. We compare four configurations: (1) \textbf{Zero-shot}: No fine-tuning, (2) \textbf{Stage 1 only}: LoRA fine-tuning with frozen encoders, (3) \textbf{Stage 2 only}: Direct encoder full fine-tuning, and (4) \textbf{Stage 1 + Stage 2}: Our complete approach.

\begin{table}[H]
\setlength{\tabcolsep}{8pt}
\renewcommand{\arraystretch}{1.15}
\begin{tabular}{lccc}
\toprule
\textbf{Training Strategy} & \textbf{mAP} & \textbf{AUC} & \textbf{acc} \\
\midrule
Zero-shot & 0.61 & 0.41 & 32.26 \\
Stage 1 only & 0.82 & 0.66 & 73.40 \\
Stage 2 only & 0.86 & 0.83 & 80.61 \\
\midrule
\textbf{Stage 1 + Stage 2 (Ours)} & \textbf{0.96} & \textbf{0.92} & \textbf{85.09} \\
\bottomrule
\end{tabular}
\caption{Ablation study on MAVOS-DD\cite{croitoru2025mavos} Open-set full scenario comparing four training configurations.}
\label{tab:ablation}
\end{table}

Table~\ref{tab:ablation} shows our complete two-stage approach achieves best performance (mAP: 0.96, AUC: 0.92, Acc: 85.09\%). Zero-shot baseline reaches 32.26\% accuracy, Stage 1 only achieves 73.40\%, and Stage 2 only reaches 80.61\%. The results demonstrate that both Stage 1 and Stage 2 are essential for optimal performance.

\subsection{Confusion Matrix Analysis}

\begin{figure}[H]
    \centering
    \includegraphics[width=0.4\textwidth, trim=0 0 0 0, clip]{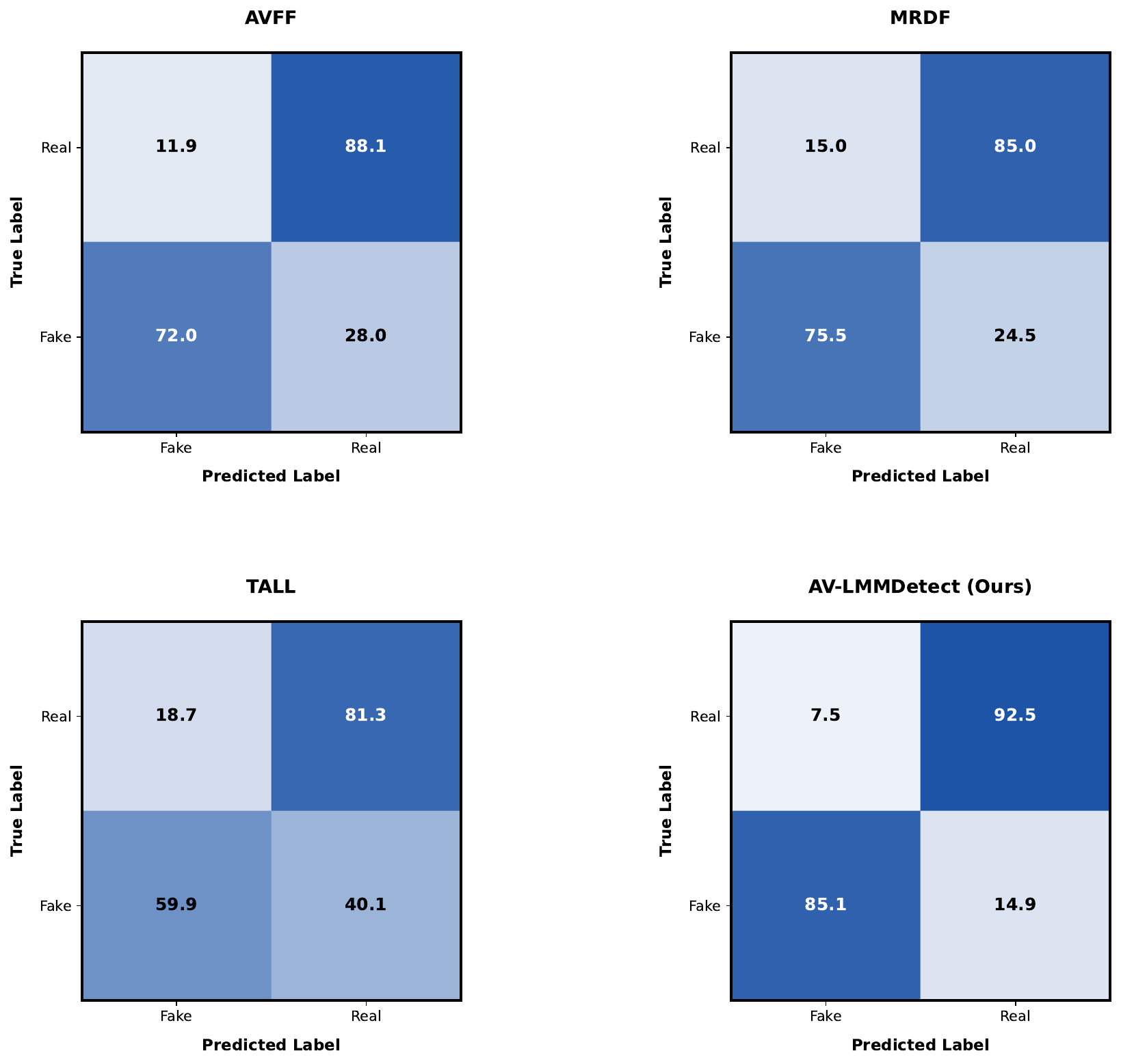}
    \caption{Confusion matrices for AVFF, MRDF, TALL, and our \textbf{\textit{AV-LMMDetect}} on MAVOS-DD Open-set full scenario.}
    \label{fig:confusion_matrices}
\end{figure}

We report the confusion matrices obtained by AVFF, MRDF, TALL, and our method for the Open-set full scenario in Figure~\ref{fig:confusion_matrices}. In this challenging scenario, AVFF shows a significant drop in its ability to detect fake videos, with 28.0\% false negatives. MRDF exhibits similar degradation with 24.5\% false negatives, although it maintains better fake detection capability than AVFF. TALL demonstrates poor performance with 40.1\% false negatives, indicating limited generalization to unseen models and languages. In contrast, our \textbf{\textit{AV-LMMDetect}} achieves superior performance with only 14.9\% false negatives and 7.5\% false positives, demonstrating robust cross-modal detection capabilities. The superior generalization ability stems from Qwen 2.5 Omni's large-scale training data and its inherent video-audio paired data structure. These observations validate that our supervised fine-tuned (SFT) approach provides better generalization across challenging open-set scenarios.

\section{Conclusion}

We introduced \textbf{\textit{AV-LMMDetect}}, the first supervised fine-tuned (SFT) large multimodal model for audio-visual deepfake detection. Our two-stage training achieves competitive results on FakeAVCeleb\cite{khalid2021fakeavceleb} and state-of-the-art performance on MAVOS-DD\cite{croitoru2025mavos}. Analyses validate our approach's effectiveness, demonstrating superior generalization across challenging open-set scenarios.

\newpage
\bibliographystyle{IEEEbib}
\bibliography{icassp_reference}

\end{document}